\documentclass[aps,pre,showpacs,amssymb,a4paper,twocolumn]{revtex4-1}%
\usepackage{amsmath}%
\usepackage{amsfonts}%
\usepackage{amssymb}%
\usepackage{graphicx}
\usepackage[english]{babel}
\usepackage{color}

\begin{document}

\title{Numerical calculation of granular entropy} 
\author{Daniel Asenjo} 
\author{Fabien Paillusson} 
\thanks{D.~Asenjo and F.~Paillusson contributed equally to this work.}
\author{Daan Frenkel} 
\affiliation{Department of Chemistry, University of
  Cambridge, Lensfield Road, Cambridge, CB2 1EW, U.K.}
\date{\today}

\begin{abstract}
  We present numerical simulations that allow us to compute the number
  of ways in which $N$ particles can pack into a given volume $V$.
  Our technique modifies the method of Xu et al. (Phys. Rev. Lett. 106,
  245502 (2011)) and outperforms existing direct enumeration methods by 
more than 200 orders of magnitude.
We use our approach to study the system size dependence of the number of
  distinct packings of a system of up to 128 poly-disperse soft disks. We show that,
  even though granular particles are distinguishable, we have to
  include a factor $1/N!$ to ensure that the entropy does not change
  when exchanging particles between systems in the same macroscopic
  state. Our simulations provide strong evidence that
  the packing entropy, when properly defined, is extensive. 
As different packings are created with unequal probabilities, it is natural to express the packing entropy  as $S=-\sum_i p_i\ln
  p_i -\ln N!$, where $p_i$ denotes the probability to generate the
  $i$-th packing.  We  can compute this quantity reliably
and it is also extensive. The granular entropy thus
  (re)defined, whilst distinct from the one proposed by Edwards (J. Phys.: Condens.Matter 2,
  SA63(1990)), does have all the properties Edwards assumed. \end{abstract}

\maketitle

\section{Introduction} 

Granular systems undergo a `jamming' transition at a density
  that is well below the highest packing density of an
  equilibrated system \cite{Zhang2009, PicaCiamarra10}. If a granular medium, such as sand, is poured into a container, it will come to rest in a mechanically stable packing. 
If the same amount of sand is, once again, poured into the same container, it will most likely come to rest in a different stable structure.
For all but the smallest systems, the total number of distinct granular packings is extremely large and it is generally assumed that the number of packings increases exponentially with system size. Below we present numerical simulations that indicate otherwise. 
 Computing the number of distinct granular packings ($\Omega(N,V)$) of $N$ particles in a volume $V=L^d$ ($d=1,2,3$ being the spacial dimensions) is not just an interesting intellectual exercise: in Edwards' theory of the flow of powders, the granular `entropy', defined as the logarithm of $\Omega$, plays a key role \cite{Edwards89, Edwards90}. Edwards' theory makes the assumption that all granular packings are equally likely at fixed packing fraction and that the logarithm of the number of packings defines a `granular entropy' that is extensive. Ever since its formulation, the validity of Edwards' hypothesis
  has been hotly debated ~\cite{Makse12, Lechenault06, Barrat2000,
    Gao09, McNamara09, Daniels13, Paillusson12, Eastham06, Wang10,
    PicaCiamarra12}. However, as direct tools to calculate the granular entropy of granular media for large systems were lacking, the debate remained inconclusive. This is because this quantity could not be computed directly for realistic, off-lattice models unless the systems were so small ($N=16$) that the packings could be counted by direct enumeration (see e.g.~\cite{Xu11, Hoy12}). Other approaches have been used to estimate granular entropy for slightly larger systems \cite{Barrat2000, Makse02, Makse12}, but no calculation to date could reliably test whether $\ln\Omega$ is extensive.  Hence, the ability to compute $\Omega$ for systems containing more than a dozen particles is, in itself, of interest.

In this letter, we estimate the entropy $\ln \Omega(N,\phi)$ of
quasi-rigid disks over a range of system sizes. To do so, we use a method where we compute the volume of basins of attraction of
individual minima on the potential energy landscape using a scheme
introduced in \cite{Xu11}. 
\section{General method}
In what follows, we will use a model that is similar to the one used
by O'Hern et al.~\cite{OHern03}. In the study of jamming of
ref.~\cite{OHern03}, particles were assumed to interact via a
(generalised) Hertzian potential and distinct granular packings were
generated by preparing the system in an ideal gas configuration and
then performing a Stillinger quench~\cite{Stillinger} to the
corresponding potential energy minimum.  In the present work we study
packings of particles that, whilst not exactly hard, approach the
hard-core limit more closely than the Hertzian potential. To this end
we consider a fluid of (polydisperse) hard disks that have, in
addition, a finite-ranged soft repulsion. The total interaction
potential $\varphi(r_{ij})$ between particles $i$ and $j$ is of the
form
\begin{equation}
\varphi(r_{ij}) = \left\lbrace \begin{array}{cc}
 +\infty , & r_{ij} < d^\text{HD}_{ij} \\ {\rm WCA}(r_{ij}-d^\text{HD}_{ij}), & d^\text{HD}_{ij} < r_{ij} < d^{S}_{ij} \\ 0 , & r_{ij} > d^{S}_{ij}
\end{array}
 \right. \label{eq4}
\end{equation}
where ${\rm WCA}(r)$ denotes the Weeks-Chandler-Andersen (WCA) potential
\cite{WCA1971}.
We assume additivity of the  hard-core radii of different particles: $d^\text{HD}_{ij}= (d^\text{HD}_{ii}+d^\text{HD}_{jj})/2$. Similarly, for the WCA range, $d_{ij}^S$ we use:
$d_{ij}^S=(d_{ii}^S+d_{jj}^S)/2$. 

In order to compute the number of distinct minima $\Omega(N,\phi)$
on the energy landscape of a system of $N$ particles at packing
fraction $\phi$, we make use of the fact that any given energy minimum can be reached by energy minimisation from a set of points that defines the basin of attraction of that minimum. 
We define the mean hyper-volume $\langle
v\rangle(N,\phi)$ of the basins of attraction of minima on this energy
landscape via:
\begin{equation}
\langle v\rangle(N,\phi) \equiv \frac{1}{\Omega(N,\phi)}\sum_{i=1}^{\Omega(N,\phi)} v_i = \frac{V_\text{acc}(N,\phi)}{\Omega(N,\phi)},\label{eq0}
\end{equation}
where $v_i$ is the hyper-volume of basin $i \in [1,\Omega(N,\phi)]$. From now on, we will refer to a basin hyper-volume as its {\it volume} and explictely refer to the box or a basin when necessary.
It is important to note that these  basins tile the accessible configuration space of the hard-core parent system. This accessible phase-space volume $V_\text{acc}(N,\phi)$ is equal to the 
configurational integral of the hard-core fluid. We can easily compute this quantity if we know the equation of state of the hard-core fluid. 
By inverting
Eq. \eqref{eq0}, we obtain $\Omega(N,\phi)=V_\text{acc}(N,\phi)/\langle v
\rangle(N,\phi)$. Hence  $\Omega(N,\phi)$ can be estimated if we know $\langle v\rangle(N,\phi)$.
The important point is that $\langle v\rangle(N,\phi)$  can be estimated by sampling a small subset of all possible jammed configurations.  
In practice, we first generate $\mathcal{N}_f$  well-equilibrated configurations of the hard-core fluid and perform a Stillinger quench on each of them. 
As explained in ref.~\cite{Xu11}, the volume $v_i(N,\phi)$ of a basin $i$ can be
calculated using thermodynamic integration from a harmonic
reference state \cite{Frenkel2001}:
\begin{equation}
  F_i(N,\phi) = F_\text{Harmonic}(N,\phi) - \frac{1}{2}\int_{0}^{k^\text{max}_i} dk \:\langle\mathbf{u}^2(k)\rangle \label{eq1}
\end{equation}
where we introduce the free energy $F_i(N,\phi) \equiv -\ln
v_i(N,\phi)$ and where $\langle\mathbf{u}^2(k)\rangle $ denotes the
canonical average of the square of the displacement vector
$\mathbf{u}$ from the lattice positions corresponding to the minimum
$i$.  The energy cost for such a displacement is $k\mathbf{u}^2/2$
if $\mathbf{u}$ is inside the basin $i$, and infinite otherwise. For $k=k^\text{max}_i$,
 $\langle\mathbf{u}^2(k)\rangle $ is estimated by static Monte Carlo sampling. For all other values of $k$ we use 
 a Markov-chain Monte Carlo sampling, combined with parallel tempering (see e.g.~ \cite{ptearl}) to speed up the convergence.
 We choose  $k^\text{max}_i$ such that  $85-95$\% of
random displacements generated from minimum $i$ are within its basin
of attraction.
\newline All allowed configurations of the equilibrated hard-sphere fluid are equally likely to be sampled, and
hence the number of quenches started in a given basin will be proportional to its volume. Because each basin has in principle a different volume from the others, it implies that sampled basins are not found with equal probability as already reported in previous studies \cite{Gao09,Xu11}. As discussed in refs. \cite{Ashwin12} and \cite{Daan13}, such non-uniform probabilities are a consequence of the protocol used to generate the jammed packings. We must therefore distinguish between $\mathcal{B}(F|N,\phi)$ and
$\mathcal{U}(F|N,\phi)$, the biased and unbiased distributions of free
energies respectively. Measuring volumes from our $\mathcal{N}_f$
minima, the raw data gives us $\mathcal{B}(F|N,\phi)$ that we can {\it
  in principle} relate to $\mathcal{U}(F|N,\phi)$ via:
\begin{equation}
  \mathcal{U}(F|N,\phi) = \mathcal{C}(N,\phi)\mathcal{B}(F|N,\phi)e^{F} 
\label{eq2}
\end{equation}
where $\mathcal{C}$ is a normalizing constant. It is then easy to see
that:
\begin{equation}
  \langle v \rangle(N,\phi) = \mathcal{C}(N,\phi) = \left[\int_{F_\text{min}}^{+\infty} dF\:\mathcal{B}(F|N,\phi)e^{F} \right]^{-1} 
\label{eq3}
\end{equation}
which is the expression that we use in what follows.

\section{System and Minimization algorithm}
We prepared a system of $N$ poly-disperse hard disks in a box of unit
size with periodic boundary conditions. We assume that the distribution of the areas of the disks is Gaussian.
We denote the mean disk area by $\overline{\mathcal{A}}$
and its standard deviation by $\sigma_{\mathcal{A}}=0.2\overline{\mathcal{A}}$. The  packing
fraction of the hard disks is $\phi_\text{HD} = N\overline{\mathcal{A}}/V_{box}
=N\overline{\mathcal{A}} $. We choose $\phi_\text{HD}$ sufficiently small to ensure that 
the hard disk fluid is not glassy and can be easily equilibrated. 
For this system, we then generate $\mathcal{N}_{f} \sim 1000$ equilibrated fluid configurations. We then switch on the soft repulsive WCA interaction such that $d_{ij}^S = (1+x/100) d_{ij}^{HS}$.  We choose $x$ such that the effective volume fraction of the  WCA particles is sufficiently high ($\phi = 0.88$) to guarantee that all energy minima of the WCA fluid are jammed {\it i.e} have non-zero energy values.

To quench the $\mathcal{N}_f$ fluid configurations into their
corresponding minimum, we use a modified version of the FIRE algorithm
\cite{Bitzek2006} that contains an extra MC-like step to prevent hard
disk overlaps during the minimization procedure. While other studies
similar to ours have used the L-BFGS algorithm \cite{Nocedal89}, we
found that the latter procedure generally produces disconnected basins
of attraction that are unsuited for the volume determination by
thermodynamic integration.  As discussed in detail in
\cite{Asenjo2013}, the FIRE algorithm requires up to four times as
many function evaluations as L-BFGS to reach a minimum, but it does
not suffer from the L-BFGS pathologies.

\section{Free energy distributions}
\begin{figure}
 \hspace{5mm} $d_{ij}^{S}/d_{ij}^\text{HD}=1.4$  \hspace{17mm}   $d_{ij}^{S}/d_{ij}^\text{HD}=1.12$\\
  \includegraphics[width=0.49\columnwidth]{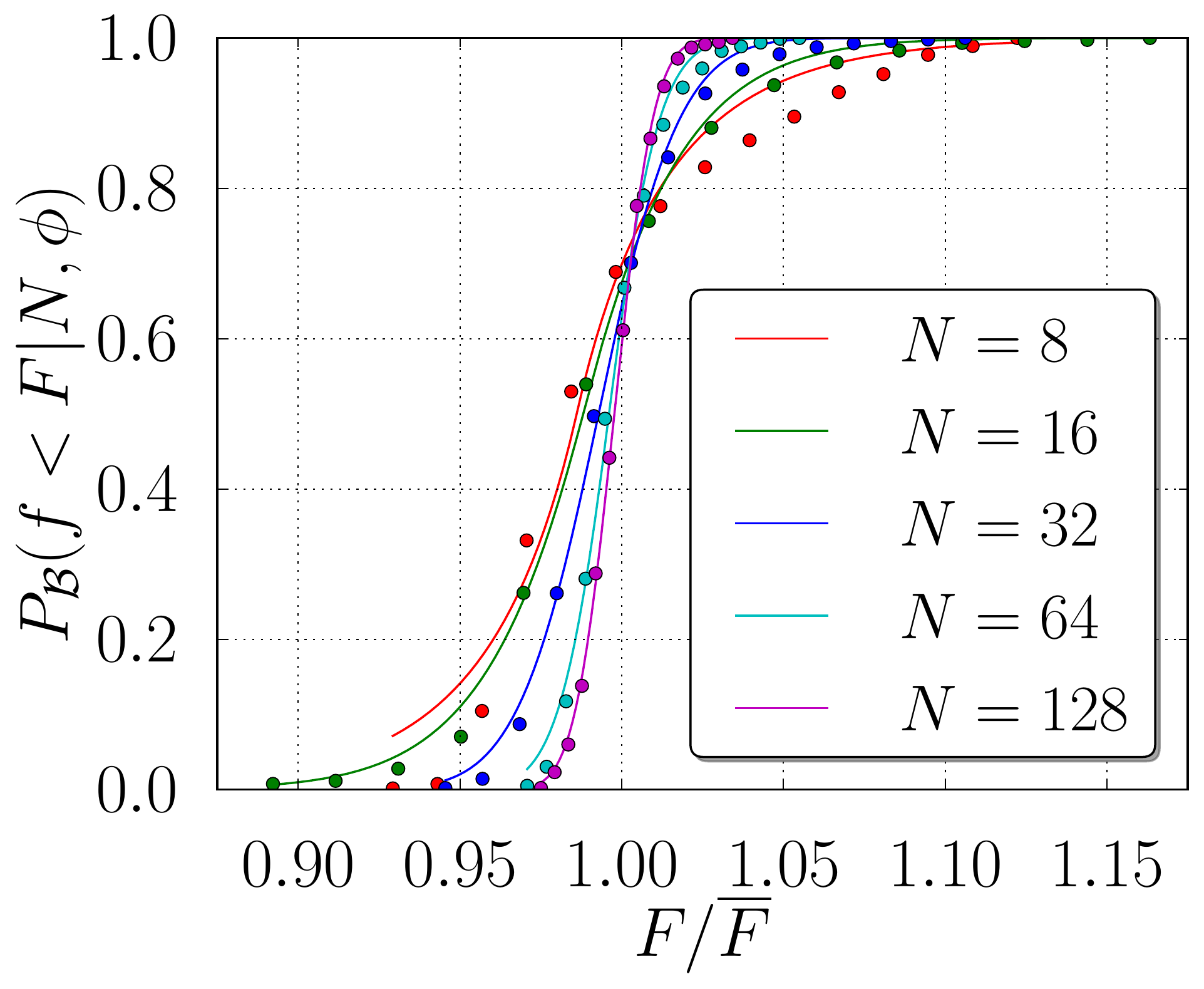} 
  \includegraphics[width=0.49\columnwidth]{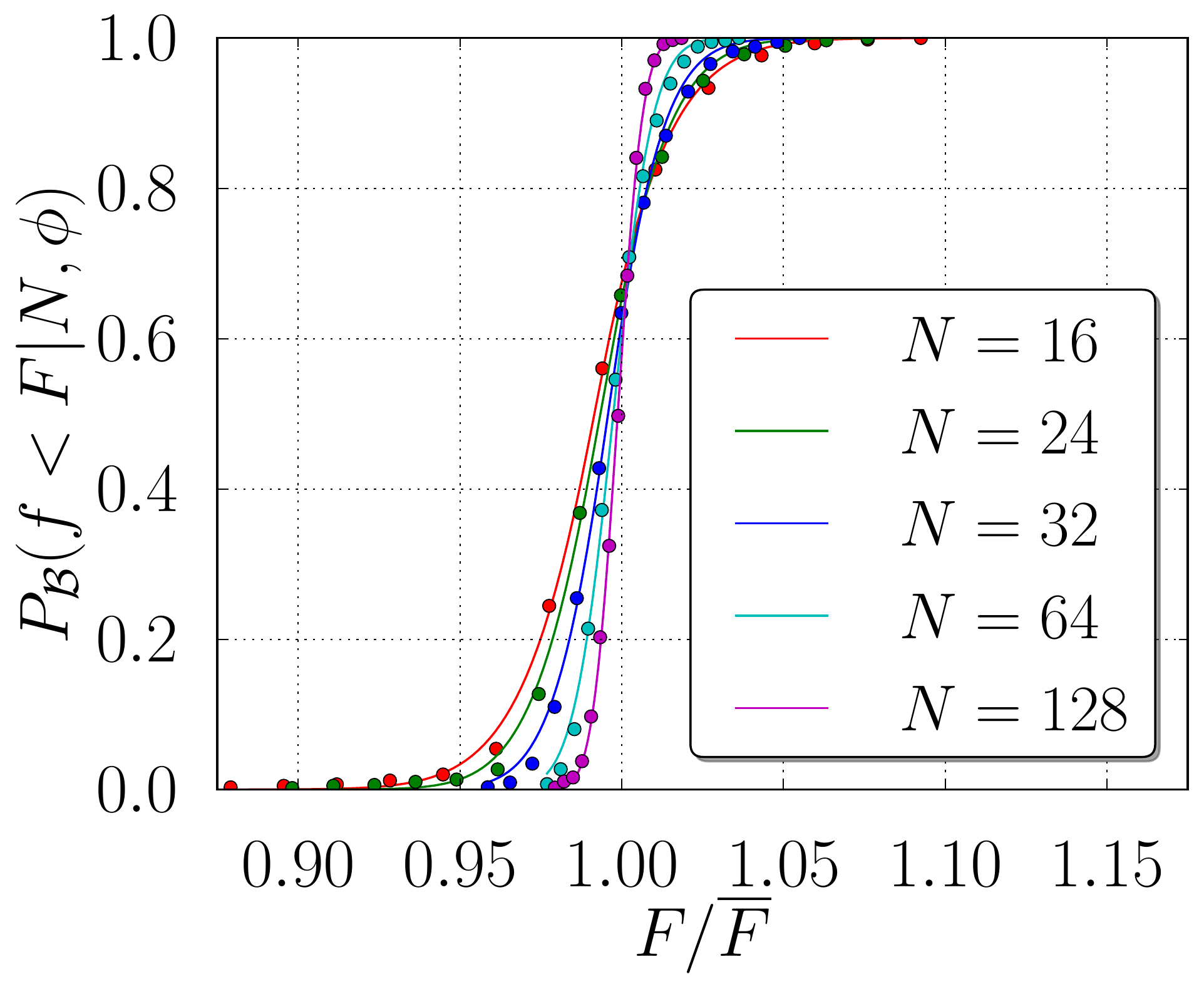}\\
  \includegraphics[width=0.49\columnwidth]{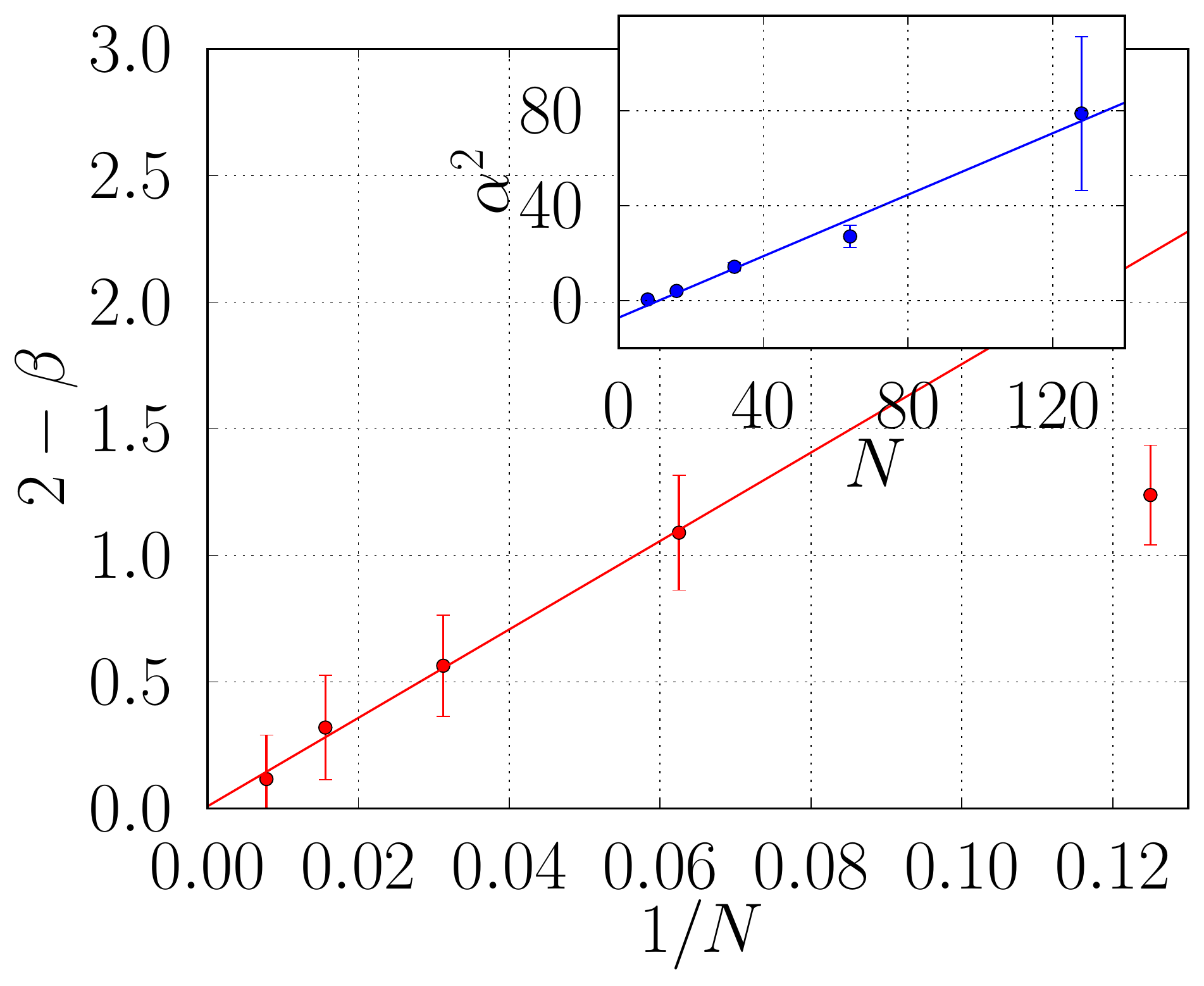} 
  \includegraphics[width=0.49\columnwidth]{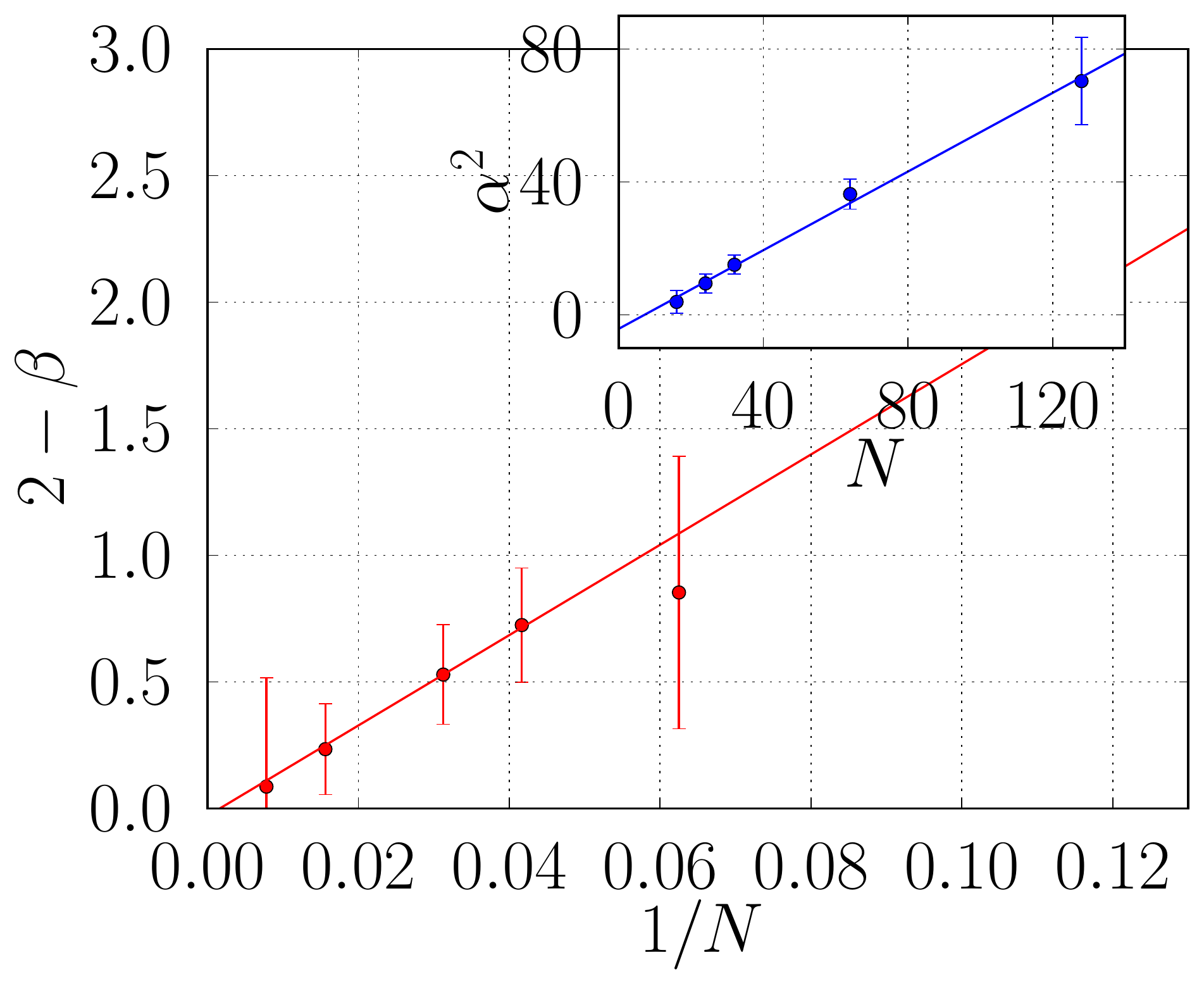} 
  \caption{\label{fig1} Properties of the biased cumulative distribution
    $\mathcal{P}_\mathcal{B}(f < F|N,\phi) \equiv \int_{-\infty}^F \:\mathcal{B}(f|N,\phi)df$ for different values of
    $d_{ij}^\text{HD}/d_{ij}^S$ and $\phi = 0.88$. The top figures
    correspond to simulation data compared to the best fitting
    generalized normal distributions from Eq. \eqref{eq5} (lines) for
    different values of N. The bottom figures show the corresponding
    values for two of the fitted parameters as a function of system
    size. $2-\beta$ is plotted as a function of the inverse system
    size $1/N$ (points) compared to a linear function. $\alpha^2$ (blue) as a function of $N$. Error bars have been obtained for all
    the parameters using the bootstrapping method \cite{bootstrap}
    with 1000 replicas.}
\end{figure}

With the improved numerical techniques, we are able to sample basin volumes for systems containing up to 128 particles, which is almost an order of magnitude larger than what was feasible before. In particular, we can obtain the basin-volume weighted distribution $\mathcal{B}(F|N,\phi)$ as a function of $N$ (see Fig. \ref{fig1}).

Our first objective is to compute the number of distinct packings. After that, we turn to the granular entropy.
We have to discuss granular entropy separately because the very fact that different basins have different volumes already implies that, in our Stillinger-quench procedure, they will not be equally populated: hence we will have to modify Edward's definition of the granular entropy from $S=\ln \Omega$ to $S^*=-\sum_i p_i\ln p_i$.  Yet, to count the number of packings we need information about the unbiased distribution of basins. We cannot {\it directly} apply Eq. \eqref{eq2} to obtain this unbiased distribution because small basin volumes are inadequately sampled. 
From Fig. \ref{fig1}, we indeed see that there is a typical spread in free
energies $\Delta F$ of at least $20$ in the sampled basins. Since we
have only $\mathcal{N}_f \sim 1000$ points in the histogram, it
implies that the most probable --- biased --- basins are about
$\mathcal{O}(10^3)$ more probable than the smallest ones. Upon unbiasing,
this ratio is multiplied by a factor $e^{-20}$ and we thus conclude
that the smallest basins, for which we have poor statistics,  are much more numerous than the large ones. 
To overcome this difficulty, we perform instead an {\it indirect} unbiasing on best fitting distributions.
Because both the biased and the true unbiased
distributions are normalisable, it implies that $\mathcal{B}(F|N,\phi)$ must decay with a functional form $e^{-F^{\nu}}$ where
$\nu > 1$. Moreover, we assume that the unbiased distribution
$\mathcal{U}$ is uni-modal - something that has been verified for small systems by direct
enumeration~\cite{Xu11}. We then fit  the observed distributions $\mathcal{B}$ with a
3-parameter Generalised Normal Distribution $p(F|\overline{F},\alpha,
\beta)$ that reads:
\begin{equation}
p(F|\overline{F},\alpha, \beta) \equiv \frac{\beta}{2\alpha \Gamma(1/\beta)}e^{-\left(\frac{|F-\overline{F}|}{\alpha} \right)^{\beta}}, \label{eq5}
\end{equation}
where $\Gamma(x)$ is the Euler Gamma function (see
Fig. \ref{fig1}).  In Eq. \eqref{eq5}, the
mean of $F$ is $\overline{F}$ and its variance is $\alpha^2
\Gamma(3/\beta)/\Gamma(1/\beta)$. In the limit $\beta\rightarrow 2$, we recover the normal distribution with width $\alpha$. 
Fig. \ref{fig1} shows the system size dependence of the best fit
parameters $\alpha^2$ and $\beta$.
We observe that $\alpha^2$ scales linearly with $N$
while $\beta$ tends to be $2$ in the large-$N$ limit. 
We stress both observations are not a priori obvious. However, the finding that 
$\mathcal{B}(F|N,\phi)$ tends to a Gaussian for large $N$ is compatible with the conjecture of refs.~\cite{Xu11} and \cite{Makse12} based on results for much smaller systems.
Using the fitted functional form for the volume distribution, we can estimate the unbiased distribution of volumes and use Eq.~\ref{eq3} to obtain an estimate for the average basin volume. 

\section{Entropy and extensivity}
To obtain an estimate of the number of packings,  we combine our information about $\langle v\rangle_{\phi,N}$ with our knowledge of the accessible volume of the parent system (i.e. the polydisperse hard-disk fluid):
\begin{equation}
- \ln V_\text{acc}(N,\phi_\text{HD}) = -\ln V^N + Nf_\text{ex}(\phi_\text{HD}) \label{eq6}
\end{equation}
where the first term on the r.h.s. stems from an ideal gas of distinguishable particles contribution where $V$ is the volume of the box (which, in our units is equal to one). $ f_\text{ex}(\phi_\text{HD})$ is the excess free energy term of the hard-disk fluid at volume fraction  $\phi_\text{HD}$. 
\begin{figure}
\includegraphics[width=0.8\columnwidth]{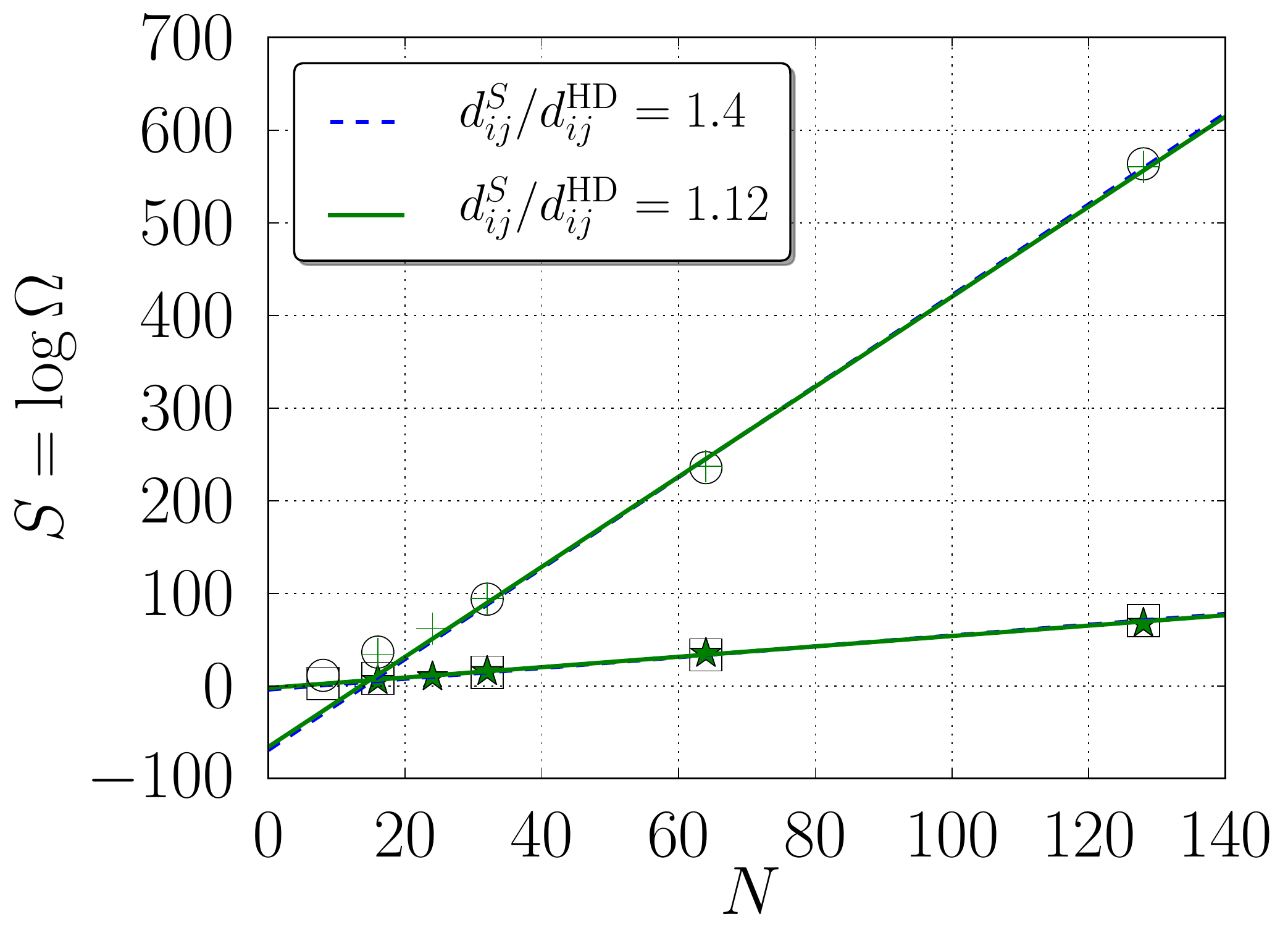}  
\caption{\label{fig2} Configurational entropy for $\phi=0.88$ and two different $d_{ij}^{S}/d_{ij}^\text{HD}$. $S(N)=\ln\Omega(N)$ (open circles and crosses for $d_{ij}^{S}/d_{ij}^\text{HD}$ = 1.4 and 1.12 respectively) comes from numerical integration of Eqs. \eqref{eq0}, \eqref{eq3} and \eqref{eq6} for different $N$. As the straight-line fits show, $S(N)$ is not extensive while $S(N)-\ln N!$ (open squares and stars for $d_{ij}^{S}/d_{ij}^\text{HD}$ = 1.4 and 1.12 respectively) appears to be extensive. We also note that the results for systems prepared with $d_{ij}^{S}/d_{ij}^\text{HD}$=1.4 and 1.12 (see text) are virtually indistinguishable.}
\end{figure}
\begin{figure}
\includegraphics[width=0.85\columnwidth]{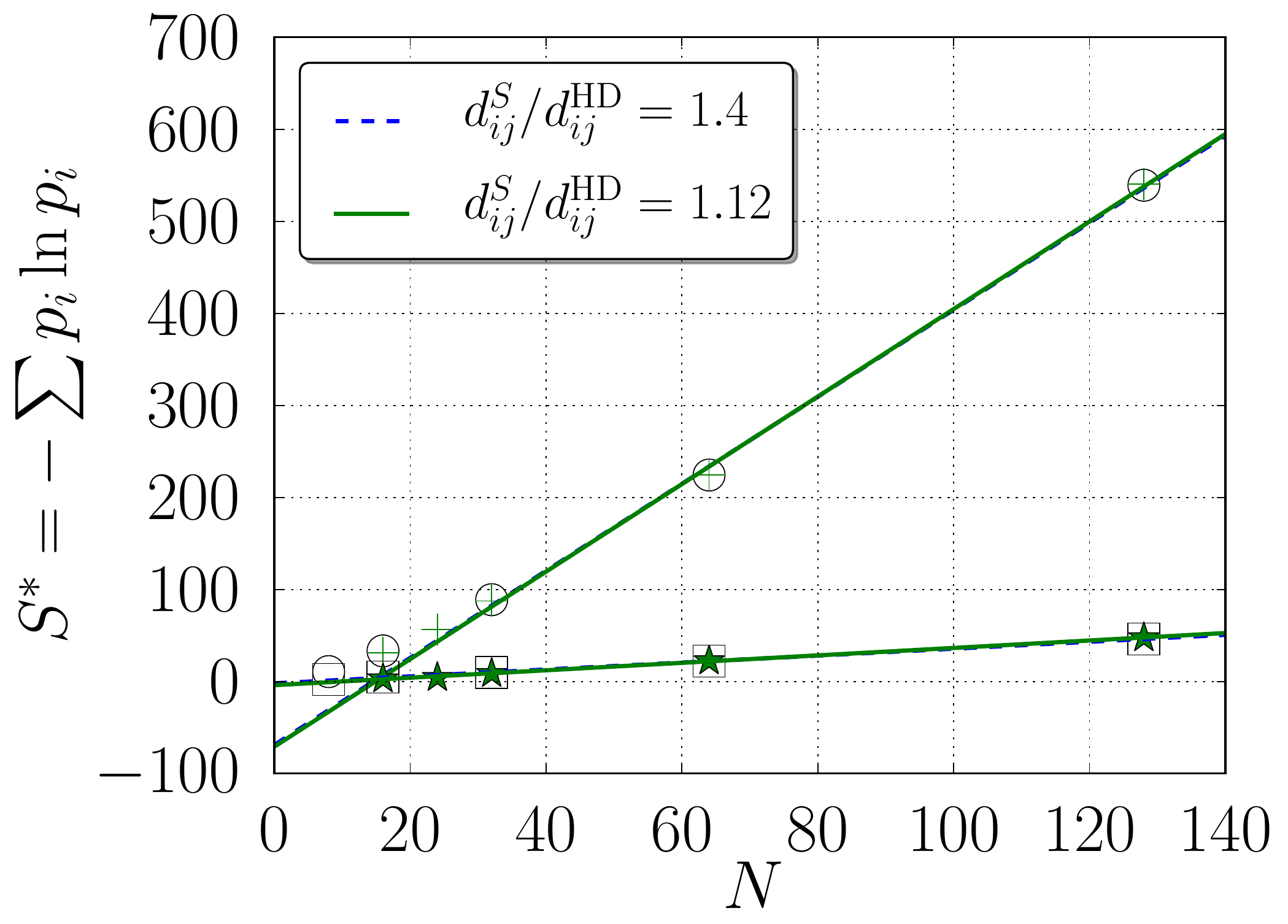} 
\caption{\label{fig3}Redfined configurational entropy for $\phi=0.88$ and two different $d_{ij}^{S}/d_{ij}^\text{HD}$. $S^*(N)=-\sum_i p_i \ln p_i$ (open circles and crosses for $d_{ij}^{S}/d_{ij}^\text{HD}$ = 1.4 and 1.12 respectively) comes from numerical integration of Eqs. \eqref{eq6} and \eqref{eq:Splnp} for different $N$. As the straight-line fits show, $S^*(N)$ is not extensive while $S^*(N)-\ln N!$ (open squares and stars for $d_{ij}^{S}/d_{ij}^\text{HD}$ = 1.4 and 1.12 respectively) appears to be extensive. We also note that the results for systems prepared with $d_{ij}^{S}/d_{ij}^\text{HD}$=1.4 and 1.12 (see text) are virtually indistinguishable.}
\end{figure}
Our estimate for $\Omega(N,\phi)$ is then obtained by combining Eqs. \eqref{eq0}, \eqref{eq3} and \eqref{eq6}. Coming back to Eq.\eqref{eq6}, since we work with a unit box, only $f_\text{ex}$ needs to be computed. We can get it directly via thermodynamic integration of $(Z(\phi)-1)/\phi$ on the interval $[0,\phi_\text{HD}]$ where $Z(\phi)\equiv P/\rho k_B T$ is the compressibility factor of the polydisperse hard-disk fluid. For polydisperse hard disks, $Z$ is well approximated by  $Z(\phi)=p Z_{m}(\phi) + \phi(1-p)/(1-\phi)$~\cite{Yuste99},  where $p \equiv
\sqrt{\langle d_\text{HD}^2\rangle}/\langle d_\text{HD}\rangle$ and $Z_m(\phi) \approx [1-2\phi-(1-2\phi_\text{max})\phi^2/\phi^2_\text{max}]^{-1}$ \cite{Yuste98}. 
Fig. \ref{fig2} shows a plot of $\ln\Omega$ versus $N$. As can be seen from the figure, the values of $\Omega$ for large $N$ are extremely large ($\mathcal{O}(10^{250})$) and could never be obtained by direct enumeration. 

In what follows, we shall make a distinction between $\ln\Omega$ and
the granular entropy. The reason is two-fold: first of all, as
different packings have different basins of attraction, they are {\em
  not} populated equally. Yet, only when all packings are equally
likely, can we expect $\ln\Omega$ to be a measure for the granular
entropy. Hence,we should use the more general expression for the
entropy of system with states that are not equally likely
\begin{equation}\label{eq:Splnp}
S^* \equiv -\sum_i p_i\ln p_i  = -\langle \ln v \rangle_{B} + \ln V_\text{acc}(N,\phi_\text{HD})
\end{equation}
The important point to note is that this expression depends only on
the basin volumes {\em as sampled} and hence requires no additional
assumptions to achieve unbiasing. Hence, we can obtain reliable
estimates for $S^*$. When we plot $S^*$ as a function of $N$
(Fig. \ref{fig3}), we note that the dependence is not very linear, in
other words: $S^*$ is not extensive. This should come as no surprise
because also in equilibrium statistical mechanics, the partition
function of a system of $N$ distinct objects is not extensive. We must
follow Gibbs's original argument 
that we should subtract $\ln N!$ from $S^*$ to ensure that two systems in identical macroscopic states can be in equilibrium under the exchange of particles, even though these particles are distinguishable in the quantum-mechanical sense.. We note that
there is much confusion in the literature on this topic, although
papers by van Kampen, Jaynes, Warren and Swendsen ~\cite{vanKampen84,Jaynes92,Warren, Swendsen06} clarify
the issue. Indeed, if we plot $S^*-\ln N!$ versus $N$ we obtain a straight line that, moreover, goes through the origin. We note that the assumption that all packings are equally
likely was an unnecessarily strong condition to construct a granular
entropy and, from that, granular thermodynamics. Edwards' hypothesis
of the existence of a meaningful granular entropy therefore survives
when the condition that all granular packings are equally likely
is dropped~\cite{Mehta}.

Our finding that the granular entropy (as defined here) is indeed
extensive (and additive) is highly significant. Extensivity of the granular entropy is crucial for a meaningful (systemsize-independent) definition of equilibrium between different granular materials.  Now that we can compute $S$,
we can start to test these theories. Of course, it would be interesting to test if
other protocols to generate jammed structures also find an extensive
granular entropy.

One interesting observation is the following: we find that the plots for
the entropy of jammed packings at $\phi=0.88$ that were generated from
two rather different parent systems (one with a short-ranged the other with
a long-ranged WCA potential) are almost on top of one another (see Fig. \ref{fig2}).
Again, this finding is not
obvious a priori. It seems to imply that effectively all minima that are
generated in the system with a low initial density are also
permissible for the high-density parent system. We do not
expect that such protocol independence of jammed structures at a given
density will hold in general. However, this finding suggests that the results that we report for soft disks  may also apply to hard particles.

 \section{Conclusion}
In this paper we have presented numerical simulations that allow us to estimate the number of distinct packings of up to 128 polydisperse hard disks under conditions were direct enumeration is utterly infeasible.  For instance, for  $N=128$ we estimate that the number of distinct states is of order $10^{250}$, a number well outside the reach of any direct enumeration scheme. 

If, in our definition of the granular entropy, we take into account that different packings are not {\em a priori} equally likely, we can use the appropriate (`canonical') form for the entropy ($S^*\equiv-\sum_i p_i\ln p_i$). This entropy can be computed accurately and without fitting to a particular form. We find that the behaviour of $S^*$ is very similar to that of $\ln\Omega$ and even the numerical values differ but little. However, the $N$-dependences of  $S^*$ (or, for that matter, of $\ln\Omega$) is distinctly non linear. 
Upon insertion of a $-\ln N!$ correction in our definition of granular entropy we obtain a quantity that {\em is} extensive.

We have thus established that the key quantity in Edwards' theory of granular media, if properly redefined,  is physically
meaningful. The observed robustness of the extensivity of the Edwards'
entropy for our system of soft, repulsive particles may explain why
experiments on soft jammed granular matter \cite{Brujic11} find good
agreement with Edwards' theory, even though in that case there is also no
reason to assume that all packings are equally likely.

\begin{acknowledgments}
We gratefully acknowledge discussions with Sam Edwards, Andrea Liu, Ning Xu, Jasna Brujic, Alberto Sicilia, Thomas Stecher, Patrick Warren and Rafi Blumenfeld.
This work has been supported by the
EPSRC grant $N^{\circ}$ EP/I000844/1. D.F. acknowledges support from
ERC Advanced Grant 227758, Wolfson Merit Award 2007/R3 of the Royal
Society of London. D.A. acknowledges support from Becas Chile
CONICYT. 
\end{acknowledgments}

\bibliographystyle{apsrev4-1}
\bibliography{biblio_granular}

\begin{thebibliography}{10}%
\makeatletter
\providecommand \@ifxundefined [1]{%
 \ifx #1\undefined \expandafter \@firstoftwo
 \else \expandafter \@secondoftwo
\fi
}%
\providecommand \@ifnum [1]{%
 \ifnum #1\expandafter \@firstoftwo
 \else \expandafter \@secondoftwo
\fi
}%
\providecommand \enquote [1]{``#1''}%
\providecommand \bibnamefont  [1]{#1}%
\providecommand \bibfnamefont [1]{#1}%
\providecommand \citenamefont [1]{#1}%
\providecommand\href[0]{\@sanitize\@href}%
\providecommand\@href[1]{\endgroup\@@startlink{#1}\endgroup\@@href}%
\providecommand\@@href[1]{#1\@@endlink}%
\providecommand \@sanitize [0]{\begingroup\catcode`\&12\catcode`\#12\relax}%
\@ifxundefined \pdfoutput {\@firstoftwo}{%
 \@ifnum{\z@=\pdfoutput}{\@firstoftwo}{\@secondoftwo}%
}{%
 \providecommand\@@startlink[1]{\leavevmode\special{html:<a href="#1">}}%
 \providecommand\@@endlink[0]{\special{html:</a>}}%
}{%
 \providecommand\@@startlink[1]{%
  \leavevmode
  \pdfstartlink
   attr{/Border[0 0 1 ]/H/I/C[0 1 1]}%
   user{/Subtype/Link/A<</Type/Action/S/URI/URI(#1)>>}%
  \relax
 }%
 \providecommand\@@endlink[0]{\pdfendlink}%
}%
\providecommand \url  [0]{\begingroup\@sanitize \@url }%
\providecommand \@url [1]{\endgroup\@href {#1}{\urlprefix}}%
\providecommand \urlprefix [0]{URL }%
\providecommand \Eprint[0]{\href }%
\@ifxundefined \urlstyle {%
  \providecommand \doi [1]{doi:\discretionary{}{}{}#1}%
}{%
  \providecommand \doi [0]{doi:\discretionary{}{}{}\begingroup
  \urlstyle{rm}\Url }%
}%
\providecommand \doibase [0]{http://dx.doi.org/}%
\providecommand \Doi[1]{\href{\doibase#1}}%
\providecommand \bibAnnote [3]{%
  \BibitemShut{#1}%
  \begin{quotation}\noindent
    \textsc{Key:}\ #2\\\textsc{Annotation:}\ #3%
  \end{quotation}%
}%
\providecommand \bibAnnoteFile [2]{%
  \IfFileExists{#2}{\bibAnnote {#1} {#2} {\input{#2}}}{}%
}%
\providecommand \typeout [0]{\immediate \write \m@ne }%
\providecommand \selectlanguage [0]{\@gobble}%
\providecommand \bibinfo [0]{\@secondoftwo}%
\providecommand \bibfield [0]{\@secondoftwo}%
\providecommand \translation [1]{[#1]}%
\providecommand \BibitemOpen[0]{}%
\providecommand \bibitemStop [0]{}%
\providecommand \bibitemNoStop [0]{.\EOS\space}%
\providecommand \EOS [0]{\spacefactor3000\relax}%
\providecommand \BibitemShut [1]{\csname bibitem#1\endcsname}%
\bibitem{Zhang2009}%
  \BibitemOpen
  \bibfield{author}{%
  \bibinfo {author} {\bibfnamefont{Z.}~\bibnamefont{Zhang}}, \bibinfo {author}
  {\bibfnamefont{N.}~\bibnamefont{Xu}}, \bibinfo {author}
  {\bibfnamefont{D.~T.}\ \bibnamefont{Chen}}, \bibinfo {author}
  {\bibfnamefont{P.}~\bibnamefont{Yunker}}, \bibinfo {author}
  {\bibfnamefont{A.~M.}\ \bibnamefont{Alsayed}}, \bibinfo {author}
  {\bibfnamefont{K.~B.}\ \bibnamefont{Aptowicz}}, \bibinfo {author}
  {\bibfnamefont{P.}~\bibnamefont{Habdas}}, \bibinfo {author}
  {\bibfnamefont{A.~J.}\ \bibnamefont{Liu}}, \bibinfo {author}
  {\bibfnamefont{S.~R.}\ \bibnamefont{Nagel}},\ and\ \bibinfo {author}
  {\bibfnamefont{A.~G.}\ \bibnamefont{Yodh}},\ }%
  \bibfield{journal}{%
  \bibinfo {journal} {Nature}\ }%
  \textbf{\bibinfo {volume} {459}},\ \bibinfo {pages} {230} (\bibinfo {year}
  {2009})%
  \bibAnnoteFile{NoStop}{Zhang2009}%
\bibitem{PicaCiamarra10}%
  \BibitemOpen
  \bibfield{author}{%
  \bibinfo {author} {\bibfnamefont{M.}~\bibnamefont{Pica~Ciamara}}, \bibinfo
  {author} {\bibfnamefont{M.}~\bibnamefont{Nicodemi}},\ and\ \bibinfo {author}
  {\bibfnamefont{A.}~\bibnamefont{Coniglio}},\ }%
  \bibfield{journal}{%
  \bibinfo {journal} {Soft Matter}\ }%
  \textbf{\bibinfo {volume} {6}},\ \bibinfo {pages} {2871} (\bibinfo {year}
  {2010})%
  \bibAnnoteFile{NoStop}{PicaCiamarra10}%
\bibitem{Edwards89}%
  \BibitemOpen
  \bibfield{author}{%
  \bibinfo {author} {\bibfnamefont{S.}~\bibnamefont{Edwards}}\ and\ \bibinfo
  {author} {\bibfnamefont{R.}~\bibnamefont{Oakeshott}},\ }%
  \bibfield{journal}{%
  \bibinfo {journal} {Physica A}\ }%
  \textbf{\bibinfo {volume} {157}},\ \bibinfo {pages} {1080} (\bibinfo {year}
  {1989})%
  \bibAnnoteFile{NoStop}{Edwards89}%
\bibitem{Edwards90}%
  \BibitemOpen
  \bibfield{author}{%
  \bibinfo {author} {\bibfnamefont{S.}~\bibnamefont{Edwards}},\ }%
  \bibfield{journal}{%
  \bibinfo {journal} {J. Phys.: Condens.Matter}\ }%
  \textbf{\bibinfo {volume} {2}},\ \bibinfo {pages} {SA63} (\bibinfo {year}
  {1990})%
  \bibAnnoteFile{NoStop}{Edwards90}%
\bibitem{Makse12}%
  \BibitemOpen
  \bibfield{author}{%
  \bibinfo {author} {\bibfnamefont{K.}~\bibnamefont{Wang}}, \bibinfo {author}
  {\bibfnamefont{C.}~\bibnamefont{Song}}, \bibinfo {author}
  {\bibfnamefont{P.}~\bibnamefont{Wang}},\ and\ \bibinfo {author}
  {\bibfnamefont{H.~A.}\ \bibnamefont{Makse}},\ }%
  \bibfield{journal}{%
  \bibinfo {journal} {Phys. Rev. E}\ }%
  \textbf{\bibinfo {volume} {86}},\ \bibinfo {pages} {011305} (\bibinfo {year}
  {2012})%
  \bibAnnoteFile{NoStop}{Makse12}%
\bibitem{Lechenault06}%
  \BibitemOpen
  \bibfield{author}{%
  \bibinfo {author} {\bibfnamefont{F.}~\bibnamefont{Lechenault}}, \bibinfo
  {author} {\bibfnamefont{F.}~\bibnamefont{da~Cruz}}, \bibinfo {author}
  {\bibfnamefont{O.}~\bibnamefont{Dauchot}},\ and\ \bibinfo {author}
  {\bibfnamefont{E.}~\bibnamefont{Bertin}},\ }%
  \bibfield{journal}{%
  \bibinfo {journal} {J. Stat. Mech.}\ }%
  \textbf{\bibinfo {volume} {2006}},\ \bibinfo {pages} {P07009} (\bibinfo
  {year} {2006})%
  \bibAnnoteFile{NoStop}{Lechenault06}%
\bibitem{Barrat2000}%
  \BibitemOpen
  \bibfield{author}{%
  \bibinfo {author} {\bibfnamefont{A.}~\bibnamefont{Barrat}}, \bibinfo {author}
  {\bibfnamefont{J.}~\bibnamefont{Kurchan}}, \bibinfo {author}
  {\bibfnamefont{V.}~\bibnamefont{Loreto}},\ and\ \bibinfo {author}
  {\bibfnamefont{M.}~\bibnamefont{Sellitto}},\ }%
  \bibfield{journal}{%
  \bibinfo {journal} {Phys. Rev. Lett.}\ }%
  \textbf{\bibinfo {volume} {85}},\ \bibinfo {pages} {5034} (\bibinfo {year}
  {2000})%
  \bibAnnoteFile{NoStop}{Barrat2000}%
\bibitem{Gao09}%
  \BibitemOpen
  \bibfield{author}{%
  \bibinfo {author} {\bibfnamefont{G.-J.}\ \bibnamefont{Gao}}, \bibinfo
  {author} {\bibfnamefont{J.}~\bibnamefont{Blawzdziewicz}}, \bibinfo {author}
  {\bibfnamefont{C.~S.}\ \bibnamefont{O'Hern}},\ and\ \bibinfo {author}
  {\bibfnamefont{M.}~\bibnamefont{Shattuck}},\ }%
  \bibfield{journal}{%
  \bibinfo {journal} {Phys. Rev. E}\ }%
  \textbf{\bibinfo {volume} {80}},\ \bibinfo {pages} {061304} (\bibinfo {year}
  {2009})%
  \bibAnnoteFile{NoStop}{Gao09}%
\bibitem{McNamara09}%
  \BibitemOpen
  \bibfield{author}{%
  \bibinfo {author} {\bibfnamefont{S.}~\bibnamefont{McNamara}}, \bibinfo
  {author} {\bibfnamefont{P.}~\bibnamefont{Richard}}, \bibinfo {author}
  {\bibfnamefont{S.}~\bibnamefont{Kiesgen~de Richter}}, \bibinfo {author}
  {\bibfnamefont{G.}~\bibnamefont{Le~Ca\"er}},\ and\ \bibinfo {author}
  {\bibfnamefont{R.}~\bibnamefont{Delannay}},\ }%
  \bibfield{journal}{%
  \bibinfo {journal} {Phys. Rev. E}\ }%
  \textbf{\bibinfo {volume} {80}},\ \bibinfo {pages} {031301} (\bibinfo {year}
  {2009})%
  \bibAnnoteFile{NoStop}{McNamara09}%
\bibitem{Daniels13}%
  \BibitemOpen
  \bibfield{author}{%
  \bibinfo {author} {\bibfnamefont{J.~G.}\ \bibnamefont{Puckett}}\ and\
  \bibinfo {author} {\bibfnamefont{K.~E.}\ \bibnamefont{Daniels}},\ }%
  \bibfield{journal}{%
  \bibinfo {journal} {Phys. Rev. Lett.}\ }%
  \textbf{\bibinfo {volume} {110}},\ \bibinfo {pages} {058001} (\bibinfo {year}
  {2013})%
  \bibAnnoteFile{NoStop}{Daniels13}%
\bibitem{Paillusson12}%
  \BibitemOpen
  \bibfield{author}{%
  \bibinfo {author} {\bibfnamefont{F.}~\bibnamefont{Paillusson}}\ and\ \bibinfo
  {author} {\bibfnamefont{D.}~\bibnamefont{Frenkel}},\ }%
  \bibfield{journal}{%
  \bibinfo {journal} {Phys. Rev. Lett.}\ }%
  \textbf{\bibinfo {volume} {109}},\ \bibinfo {pages} {208001} (\bibinfo {year}
  {2012})%
  \bibAnnoteFile{NoStop}{Paillusson12}%
\bibitem{Eastham06}%
  \BibitemOpen
  \bibfield{author}{%
  \bibinfo {author} {\bibfnamefont{P.}~\bibnamefont{Eastham}}, \bibinfo
  {author} {\bibfnamefont{R.}~\bibnamefont{Blythe}}, \bibinfo {author}
  {\bibfnamefont{A.}~\bibnamefont{Bray}},\ and\ \bibinfo {author}
  {\bibfnamefont{M.~A.}\ \bibnamefont{Moore}},\ }%
  \bibfield{journal}{%
  \bibinfo {journal} {Phys. Rev. B}\ }%
  \textbf{\bibinfo {volume} {74}},\ \bibinfo {pages} {020406} (\bibinfo {year}
  {2006})%
  \bibAnnoteFile{NoStop}{Eastham06}%
\bibitem{Wang10}%
  \BibitemOpen
  \bibfield{author}{%
  \bibinfo {author} {\bibfnamefont{K.}~\bibnamefont{Wang}}, \bibinfo {author}
  {\bibfnamefont{C.}~\bibnamefont{Song}}, \bibinfo {author}
  {\bibfnamefont{P.}~\bibnamefont{Wang}},\ and\ \bibinfo {author}
  {\bibfnamefont{H.~A.}\ \bibnamefont{Makse}},\ }%
  \bibfield{journal}{%
  \bibinfo {journal} {Eur. Phys. Lett.}\ }%
  \textbf{\bibinfo {volume} {91}},\ \bibinfo {pages} {68001} (\bibinfo {year}
  {2010})%
  \bibAnnoteFile{NoStop}{Wang10}%
\bibitem{PicaCiamarra12}%
  \BibitemOpen
  \bibfield{author}{%
  \bibinfo {author} {\bibfnamefont{M.}~\bibnamefont{Pica~Ciamarra}}, \bibinfo
  {author} {\bibfnamefont{P.}~\bibnamefont{Richard}}, \bibinfo {author}
  {\bibfnamefont{M.}~\bibnamefont{Schr\"oter}},\ and\ \bibinfo {author}
  {\bibfnamefont{B.}~\bibnamefont{Tighe}},\ }%
  \bibfield{journal}{%
  \bibinfo {journal} {Soft Matter}\ }%
  \textbf{\bibinfo {volume} {8}},\ \bibinfo {pages} {9731} (\bibinfo {year}
  {2012})%
  \bibAnnoteFile{NoStop}{PicaCiamarra12}%
\bibitem{Xu11}%
  \BibitemOpen
  \bibfield{author}{%
  \bibinfo {author} {\bibfnamefont{N.}~\bibnamefont{Xu}}, \bibinfo {author}
  {\bibfnamefont{D.}~\bibnamefont{Frenkel}},\ and\ \bibinfo {author}
  {\bibfnamefont{A.~J.}\ \bibnamefont{Liu}},\ }%
  \bibfield{journal}{%
  \bibinfo {journal} {Phys. Rev. Lett.}\ }%
  \textbf{\bibinfo {volume} {106}},\ \bibinfo {pages} {245502} (\bibinfo {year}
  {2011})%
  \bibAnnoteFile{NoStop}{Xu11}%
\bibitem{Hoy12}%
  \BibitemOpen
  \bibfield{author}{%
  \bibinfo {author} {\bibfnamefont{R.~S.}\ \bibnamefont{Hoy}}, \bibinfo
  {author} {\bibfnamefont{J.}~\bibnamefont{Harwayne-Gidansky}},\ and\ \bibinfo
  {author} {\bibfnamefont{C.~S.}\ \bibnamefont{O'Hern}},\ }%
  \bibfield{journal}{%
  \bibinfo {journal} {Phys. Rev. E}\ }%
  \textbf{\bibinfo {volume} {85}},\ \bibinfo {pages} {051403} (\bibinfo {year}
  {2012})%
  \bibAnnoteFile{NoStop}{Hoy12}%
\bibitem{Makse02}%
  \BibitemOpen
  \bibfield{author}{%
  \bibinfo {author} {\bibfnamefont{H.~A.}\ \bibnamefont{Makse}}\ and\ \bibinfo
  {author} {\bibfnamefont{K.}~\bibnamefont{J.}},\ }%
  \bibfield{journal}{%
  \bibinfo {journal} {Nature}\ }%
  \textbf{\bibinfo {volume} {415}},\ \bibinfo {pages} {614} (\bibinfo {year}
  {2002})%
  \bibAnnoteFile{NoStop}{Makse02}%
\bibitem{OHern03}%
  \BibitemOpen
  \bibfield{author}{%
  \bibinfo {author} {\bibfnamefont{C.}~\bibnamefont{O'Hern}}, \bibinfo {author}
  {\bibfnamefont{L.}~\bibnamefont{Silbert}}, \bibinfo {author}
  {\bibfnamefont{A.}~\bibnamefont{Liu}},\ and\ \bibinfo {author}
  {\bibfnamefont{S.}~\bibnamefont{Nagel}},\ }%
  \bibfield{journal}{%
  \bibinfo {journal} {Phys. Rev. E}\ }%
  \textbf{\bibinfo {volume} {68}},\ \bibinfo {pages} {011306} (\bibinfo {year}
  {2003})%
  \bibAnnoteFile{NoStop}{OHern03}%
\bibitem{Stillinger}%
  \BibitemOpen
  \bibfield{author}{%
  \bibinfo {author} {\bibfnamefont{F.~H.}\ \bibnamefont{Stillinger}}\ and\
  \bibinfo {author} {\bibfnamefont{T.~A.}\ \bibnamefont{Weber}},\ }%
  \bibfield{journal}{%
  \bibinfo {journal} {Phys. Rev. A}\ }%
  \textbf{\bibinfo {volume} {28}},\ \bibinfo {pages} {2408} (\bibinfo {month}
  {Oct}\ \bibinfo {year} {1983})%
  \bibAnnoteFile{NoStop}{Stillinger}%
\bibitem{WCA1971}%
  \BibitemOpen
  \bibfield{author}{%
  \bibinfo {author} {\bibfnamefont{J.~D.}\ \bibnamefont{Weeks}}, \bibinfo
  {author} {\bibfnamefont{D.}~\bibnamefont{Chandler}},\ and\ \bibinfo {author}
  {\bibfnamefont{H.~C.}\ \bibnamefont{Andersen}},\ }%
  \bibfield{journal}{%
  \bibinfo {journal} {J. Chem. Phys.}\ }%
  \textbf{\bibinfo {volume} {54}},\ \bibinfo {pages} {5237} (\bibinfo {year}
  {1971})%
  \bibAnnoteFile{NoStop}{WCA1971}%
\bibitem{Frenkel2001}%
  \BibitemOpen
  \bibfield{author}{%
  \bibinfo {author} {\bibfnamefont{D.}~\bibnamefont{Frenkel}}\ and\ \bibinfo
  {author} {\bibfnamefont{B.}~\bibnamefont{Smit}},\ }%
  \emph{\bibinfo {title} {{Understanding Molecular Simulation: From Algorithms
  to Applications}}},\ \bibinfo {edition} {2nd}\ ed.\ (\bibinfo {publisher}
  {Academic Press},\ \bibinfo {year} {2001})\ ISBN \bibinfo {isbn}
  {0122673514}%
  \bibAnnoteFile{NoStop}{Frenkel2001}%
\bibitem{ptearl}%
  \BibitemOpen
  \bibfield{author}{%
  \bibinfo {author} {\bibfnamefont{D.~J.}\ \bibnamefont{Earl}}\ and\ \bibinfo
  {author} {\bibfnamefont{M.~W.}\ \bibnamefont{Deem}},\ }%
  \bibfield{journal}{%
  \bibinfo {journal} {Phys. Chem. Chem. Phys.}\ }%
  \textbf{\bibinfo {volume} {7}},\ \bibinfo {pages} {3910} (\bibinfo {year}
  {2005}),\ ISSN \bibinfo {issn} {1463-9076}%
  \bibAnnoteFile{NoStop}{ptearl}%
\bibitem{Ashwin12}%
  \BibitemOpen
  \bibfield{author}{%
  \bibinfo {author} {\bibfnamefont{S.~S.}\ \bibnamefont{Ashwin}}, \bibinfo
  {author} {\bibfnamefont{J.}~\bibnamefont{Blawzdziewicz}}, \bibinfo {author}
  {\bibfnamefont{C.~S.}\ \bibnamefont{O'~Hern}},\ and\ \bibinfo {author}
  {\bibfnamefont{M.~D.}\ \bibnamefont{Shattuck}},\ }%
  \bibfield{journal}{%
  \bibinfo {journal} {Phys. Rev. E}\ }%
  \textbf{\bibinfo {volume} {85}},\ \bibinfo {pages} {061307} (\bibinfo {year}
  {2012})%
  \bibAnnoteFile{NoStop}{Ashwin12}%
\bibitem{Daan13}%
  \BibitemOpen
  \bibfield{author}{%
  \bibinfo {author} {\bibfnamefont{D.}~\bibnamefont{Frenkel}}, \bibinfo
  {author} {\bibfnamefont{D.}~\bibnamefont{Asenjo}},\ and\ \bibinfo {author}
  {\bibfnamefont{F.}~\bibnamefont{Paillusson}},\ }%
  \bibfield{journal}{%
  \bibinfo {journal} {Mol. Phys.}\ }%
  \textbf{\bibinfo {volume} {111}},\ \bibinfo {pages} {3641} (\bibinfo {year}
  {2013})%
  \bibAnnoteFile{NoStop}{Daan13}%
\bibitem{Bitzek2006}%
  \BibitemOpen
  \bibfield{author}{%
  \bibinfo {author} {\bibfnamefont{E.}~\bibnamefont{Bitzek}}, \bibinfo {author}
  {\bibfnamefont{P.}~\bibnamefont{Koskinen}}, \bibinfo {author}
  {\bibfnamefont{F.}~\bibnamefont{G\"ahler}}, \bibinfo {author}
  {\bibfnamefont{M.}~\bibnamefont{Moseler}},\ and\ \bibinfo {author}
  {\bibfnamefont{P.}~\bibnamefont{Gumbsch}},\ }%
  \bibfield{journal}{%
  \bibinfo {journal} {Phys. Rev. Lett.}\ }%
  \textbf{\bibinfo {volume} {97}},\ \bibinfo {pages} {170201} (\bibinfo {month}
  {Oct}\ \bibinfo {year} {2006})%
  \bibAnnoteFile{NoStop}{Bitzek2006}%
\bibitem{Nocedal89}%
  \BibitemOpen
  \bibfield{author}{%
  \bibinfo {author} {\bibfnamefont{D.~C.}\ \bibnamefont{Liu}}\ and\ \bibinfo
  {author} {\bibfnamefont{J.}~\bibnamefont{Nocedal}},\ }%
  \bibfield{journal}{%
  \bibinfo {journal} {Math. Prog.}\ }%
  \textbf{\bibinfo {volume} {45}},\ \bibinfo {pages} {503} (\bibinfo {year}
  {1989}),\ ISSN \bibinfo {issn} {0025-5610}%
  \bibAnnoteFile{NoStop}{Nocedal89}%
\bibitem{Asenjo2013}%
  \BibitemOpen
  \bibfield{author}{%
  \bibinfo {author} {\bibfnamefont{D.}~\bibnamefont{Asenjo}}, \bibinfo {author}
  {\bibfnamefont{J.~D.}\ \bibnamefont{Stevenson}}, \bibinfo {author}
  {\bibfnamefont{D.~J.}\ \bibnamefont{Wales}},\ and\ \bibinfo {author}
  {\bibfnamefont{D.}~\bibnamefont{Frenkel}},\ }%
  \bibfield{journal}{%
  \bibinfo {journal} {J. Phys. Chem. B}}%
   (\bibinfo {year} {2013}),\ \doi{\bibinfo {doi} {10.1021/jp312457a}}%
  \bibAnnoteFile{NoStop}{Asenjo2013}%
\bibitem{bootstrap}%
  \BibitemOpen
  \bibfield{author}{%
  \bibinfo {author} {\bibfnamefont{B.}~\bibnamefont{Efron}}\ and\ \bibinfo
  {author} {\bibfnamefont{R.}~\bibnamefont{Tibshirani}},\ }%
  \emph{\bibinfo {title} {An Introduction to the Bootstrap}},\ Monographs on
  statistics and applied probabilities\ (\bibinfo {publisher} {Chapman \&
  Hall/CRC},\ \bibinfo {year} {1993})%
  \bibAnnoteFile{NoStop}{bootstrap}%
\bibitem{Yuste99}%
  \BibitemOpen
  \bibfield{author}{%
  \bibinfo {author} {\bibfnamefont{A.}~\bibnamefont{Santos}}, \bibinfo {author}
  {\bibfnamefont{S.~B.}\ \bibnamefont{Yuste}},\ and\ \bibinfo {author}
  {\bibfnamefont{M.}~\bibnamefont{L{\`o}pez~de Haro}},\ }%
  \bibfield{journal}{%
  \bibinfo {journal} {Mol. Phys.}\ }%
  \textbf{\bibinfo {volume} {96}},\ \bibinfo {pages} {1} (\bibinfo {year}
  {1999})%
  \bibAnnoteFile{NoStop}{Yuste99}%
\bibitem{Yuste98}%
  \BibitemOpen
  \bibfield{author}{%
  \bibinfo {author} {\bibfnamefont{M.}~\bibnamefont{L{\`o}pez~de Haro}},
  \bibinfo {author} {\bibfnamefont{A.}~\bibnamefont{Santos}},\ and\ \bibinfo
  {author} {\bibfnamefont{S.~B.}\ \bibnamefont{Yuste}},\ }%
  \bibfield{journal}{%
  \bibinfo {journal} {Eur. J. Phys.}\ }%
  \textbf{\bibinfo {volume} {19}},\ \bibinfo {pages} {281} (\bibinfo {year}
  {1998})%
  \bibAnnoteFile{NoStop}{Yuste98}%
\bibitem{vanKampen84}%
  \BibitemOpen
  \bibfield{author}{%
  \bibinfo {author} {\bibfnamefont{N.~G.}\ \bibnamefont{van Kampen}},\ }%
  in\ \emph{\bibinfo {booktitle} {Essays in Theoretical Physics: in Honor of
  Dirk ter Haar}},\ \bibinfo {editor} {edited by\ \bibinfo {editor}
  {\bibfnamefont{W.}~\bibnamefont{Parry}}}\ (\bibinfo {publisher} {Oxford:
  Pergamon},\ \bibinfo {year} {1984})%
  \bibAnnoteFile{NoStop}{vanKampen84}%
\bibitem{Jaynes92}%
  \BibitemOpen
  \bibfield{author}{%
  \bibinfo {author} {\bibfnamefont{E.~T.}\ \bibnamefont{Jaynes}},\ }%
  in\ \emph{\bibinfo {booktitle} {Maximum Entropy and Bayesian Methods}},\
  \bibinfo {editor} {edited by\ \bibinfo {editor}
  {\bibfnamefont{C.}~\bibnamefont{Smith}}, \bibinfo {editor}
  {\bibfnamefont{G.}~\bibnamefont{Erickson}},\ and\ \bibinfo {editor}
  {\bibfnamefont{P.}~\bibnamefont{Neudorfer}}}\ (\bibinfo {publisher} {Kluwer
  Academic},\ \bibinfo {year} {1992})\ pp.\ \bibinfo {pages} {1--22}%
  \bibAnnoteFile{NoStop}{Jaynes92}%
\bibitem{Warren}%
  \BibitemOpen
  \bibfield{author}{%
  \bibinfo {author} {\bibfnamefont{P.~B.}\ \bibnamefont{Warren}},\ }%
  \bibfield{journal}{%
  \bibinfo {journal} {Phys. Rev. Lett.}\ }%
  \textbf{\bibinfo {volume} {80}},\ \bibinfo {pages} {1369} (\bibinfo {year}
  {1998})%
  \bibAnnoteFile{NoStop}{Warren}%
\bibitem{Swendsen06}%
  \BibitemOpen
  \bibfield{author}{%
  \bibinfo {author} {\bibfnamefont{R.~H.}\ \bibnamefont{Swendsen}},\ }%
  \bibfield{journal}{%
  \bibinfo {journal} {Am. J. Phys.}\ }%
  \textbf{\bibinfo {volume} {74}},\ \bibinfo {pages} {187} (\bibinfo {year}
  {2006})%
  \bibAnnoteFile{NoStop}{Swendsen06}%
\bibitem{Mehta}%
  \BibitemOpen
  \bibfield{author}{%
  \bibinfo {author} {\bibfnamefont{A.}~\bibnamefont{Mehta}},\ }%
  \emph{\bibinfo {title} {Granular physics}}\ (\bibinfo {publisher} {Cambridge
  University Press},\ \bibinfo {year} {2007})%
  \bibAnnoteFile{NoStop}{Mehta}%
\bibitem{Brujic11}%
  \BibitemOpen
  \bibfield{author}{%
  \bibinfo {author} {\bibfnamefont{I.}~\bibnamefont{Jorjadze}}, \bibinfo
  {author} {\bibfnamefont{L.-L.}\ \bibnamefont{Pontani}}, \bibinfo {author}
  {\bibfnamefont{K.~A.}\ \bibnamefont{Newhall}},\ and\ \bibinfo {author}
  {\bibnamefont{Bruji\'c}},\ }%
  \bibfield{journal}{%
  \bibinfo {journal} {Proc. Nati. Acad. Sci.}\ }%
  \textbf{\bibinfo {volume} {108}},\ \bibinfo {pages} {4286} (\bibinfo {year}
  {2011})%
  \bibAnnoteFile{NoStop}{Brujic11}%
\end{thebibliography}%

\end{document}